\documentclass[12pt,a4paper]{article}
\usepackage{graphicx}
\usepackage{amssymb}

\begin{document}

\title{{\large \bf Moduli-Space Approximation for BPS Brane-Worlds}}

\author{Gonzalo A. Palma and Anne-Christine Davis}

\begin{centering}
\bigskip
{\LARGE \bf Moduli-Space Approximation for BPS Brane-Worlds} \\
\bigskip
{\large \bf Gonzalo A. Palma and Anne-Christine Davis} \\
\bigskip \noindent \it
Department of Applied Mathematics and Theoretical Physics,\\
Center for Mathematical Sciences, University
of Cambridge,\\ Wilberforce Road, Cambridge CB3 0WA,\\
United Kingdom.\\
\end{centering}

\vspace{1cm} \baselineskip = 22pt

\date{July 2004}

\begin{center}
{\Large \bf Abstract}
\end{center}

{\bf
We develop the moduli-space approximation for the low energy
regime of BPS-branes with a bulk scalar field to obtain an
effective four-dimensional action describing the system.
An arbitrary BPS potential is used and account is taken of the presence of
matter in the branes and small supersymmetry breaking
terms. The resulting effective theory is a bi-scalar tensor theory
of gravity. In this theory, the scalar degrees of freedom can be
stabilized naturally without the introduction of additional
mechanisms other than the appropriate BPS potential. We place observational
constraints on the shape of the potential and the global configuration
of branes.
}

\newpage


Brane-worlds scenarios are an interesting theoretical possibility
to address many questions and problems of low and high energy
physics \cite{reviews}. In these models the construction of four
dimensional effective theories has proved particularly useful
to describe the physics of branes at the low energy regime. It has
been shown \cite{scalar-tensor} that these effective theories have
much in common with multi-scalar-tensor theories of gravity, where
the inter-brane distance plays the role of a scalar degree of
freedom. This is the case, for example, of the Randall-Sundrum
model \cite{RS} where the radion field emerges in the four
dimensional description.

In this letter we construct the moduli-space approximation for the
low energy regime of a general BPS brane-world model, which has been
motivated as a supersymmetric extension of the Randall--Sundrum
model \cite{SUSY} (see \cite{pheno} for phenomenological
motivations). The model consists of a five-dimensional bulk space with a
scalar field $\psi$, bounded by two branes, $\sigma_{1}$ and
$\sigma_{2}$. The main property of this system is a special
boundary condition that holds between the branes and the bulk
fields, which allows the branes to be located anywhere in the
background without obstruction. In particular, a special relation
exists between the scalar field potential $U(\psi)$, defined in
the bulk, and the brane tensions $U_{B} (\psi^{1})$ and $U_{B}
(\psi^{2})$ defined at the position of the branes. This is the BPS
condition. Due to the difficulties of this setup, in previous
works, only the special case of dilatonic branes have been
considered, where the BPS potential has the form $U_{B} \propto
e^{\alpha \psi}$. Here we consider an arbitrary potential $U_{B}$.

\begin{figure}[ht]
\begin{center}
\includegraphics[width=0.6\textwidth]{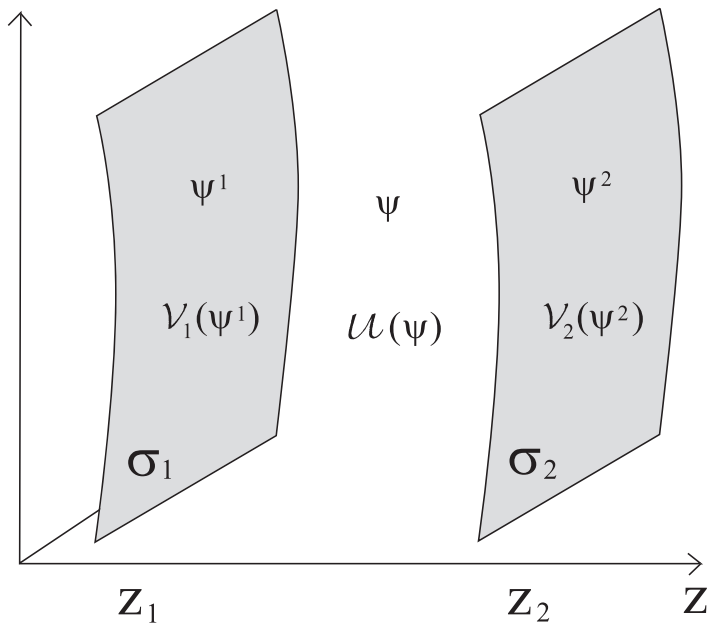}
\caption[Basic configuration]{Schematic representation of the
5-dimensional brane configuration. In the bulk there is a scalar
field $\psi$ with a bulk potential $\mathcal{U}(\psi)$.
Additionally, the bulk-space is bounded by branes, $\sigma_{1}$
and $\sigma_{2}$, with tensions given by
$\mathcal{V}_{1}(\psi^{1})$ and $\mathcal{V}_{2}(\psi^{2})$
respectively, where $\psi^{1}$ and $\psi^{2}$ are the boundary
values of $\psi$.} \label{F1}
\end{center}
\end{figure}

We now introduce the system in more detail. Let us consider a
5-dimensional manifold $M$ with the special topology $M = \sigma
\times S^{1}/ Z_{2}$, where $\sigma$ is a fixed 4-dimensional
lorentzian manifold without boundaries and $S^{1}/Z_{2}$ is the
orbifold constructed from the 1-dimensional circle with points
identified through a $Z_{2}$-symmetry. The manifold $M$ is bounded
by two branes located at the fixed points of $S^{1}/Z_{2}$. Let us
denote the brane-surfaces by $\sigma_{1}$ and $\sigma_{2}$
respectively and the space bounded by the branes as the bulk space.
In our model there is a bulk
scalar field $\psi$ living in $M$ with boundary values, $\psi^{1}$
and $\psi^{2}$, at the branes, with bulk potential $\mathcal{U}(\psi)$
and brane tensions $\mathcal{V}_{1}(\psi^{1})$ and $\mathcal{V}_{2}(\psi^{2})$
(which are potentials for the boundary values $\psi^{1}$ and $\psi^{2}$).
Additionally, we will consider the existence of matter fields
$\Psi_{1}$ and $\Psi_{2}$ confined to the branes. Figure \ref{F1}
shows a schematic representation of the present configuration.

Given the present topology, it is appropriate to introduce
foliations with a coordinate system $x^{\mu}$ describing $\sigma$
(as well as the branes $\sigma_{1}$ and $\sigma_{2}$) where
$\mu=0,\ldots,3$. Additionally, we can introduce a coordinate $z$
describing the $S^{1}/Z_{2}$ orbifold and parameterizing the
foliations. With this decomposition the following form of the line
element can be used to describe $M$ (the gaussian normal
coordinate system):
\begin{eqnarray}
ds^{2} = N^{2} dz^{2}  + g_{\mu \nu} dx^{\mu} dx^{\nu} \! .
\end{eqnarray}
Here $N$ is the lapse function for the extra dimensional
coordinate $z$ and, therefore, it can be defined up to a gauge
choice. Additionally, $g_{\mu \nu}$ is the pullback of the induced
metric on the 4--dimensional foliations, with the $(-,+,+,+)$
signature. The branes, $\sigma_{1}$ and $\sigma_{2}$, are located
at the fixed points of the $S^{1}/Z_{2}$ orbifold, denoted by $z =
z_{1}$ and $z = z_{2}$. Without loss of generality, we take $z_{1}
< z_{2}$. The total action of the system is
\begin{eqnarray} \label{eq: Tot-Act}
S_{\mathrm{tot}} = S_{\mathrm{G}} + S_{\mathrm{\psi}} +
S_{\mathrm{BR}},
\end{eqnarray}
where $S_{\mathrm{G}}$ is the action describing the pure
gravitational part and is given by $S_{\mathrm{G}} =
S_{\mathrm{EH}} + S_{\mathrm{GH}}$, with $S_{\mathrm{EH}}$ the
Einstein--Hilbert action and $S_{\mathrm{GH}}$ the
Gibbons--Hawking boundary terms. In the present parameterization:
\begin{eqnarray} \label{eq: Sg}
S_{\mathrm{G}} = \frac{1}{2 \kappa_{5}^{2}} \int_{M} \!\!\!\! dz
\, d^{\, 4} x \sqrt{-g} \, N \big( R - K_{\mu \nu} K^{\mu \nu} +
K^{2}  \big),
\end{eqnarray}
where $R$ is the four-dimensional Ricci scalar constructed from
$g_{\mu \nu}$, and $\kappa_{5}^{2} = 8 \pi G_{5}$, with $G_{5}$
the five-dimensional Newton's constant. Additionally,
\hbox{$K_{\mu \nu} = g_{\mu \nu}' / 2N$} is the extrinsic
curvature of the foliations and $K$ its trace (the prime denotes
derivatives in terms of $z$, that is $' = \partial_{z}$, and
covariant derivatives, $\nabla_{\mu}$, are constructed from the
induced metric $g_{\mu \nu}$ in the standard way). The action for
the bulk scalar field, $S_{\psi}$, can be written in the form
\begin{eqnarray}
S_{\psi} &=& - \frac{3}{8 \kappa_{5}^{2}} \int_{M} \!\!\!\! dz \,
d^{\, 4} x \sqrt{-g} \, N \big[ (\psi'/ N )^{2}
+ (\partial \psi)^{2}   + \mathcal{U}(\psi) \big] + S_{\psi}^{\,1} + S_{\psi}^{\,2} ,
\end{eqnarray}
where $S_{\psi}^{\,1}$ and  $S_{\psi}^{\,2}$ are boundary terms
given by
\begin{eqnarray} \label{eq: brane-tensions}
S_{\psi}^{\,1} &=& - \frac{3}{2 \kappa_{5}^{2}} \int_{\sigma_{1}}
\!\!\! d^{4} x \sqrt{-g} \, \mathcal{V}_{1}(\psi^{1}), \\
S_{\psi}^{\,2} &=& + \frac{3}{2 \kappa_{5}^{2}} \int_{\sigma_{2}}
\!\!\! d^{4} x \sqrt{-g} \, \mathcal{V}_{2}(\psi^{2}),
\end{eqnarray}
at the respective positions, $z_{1}$ and $z_{2}$ and
$\mathcal{U}(\psi)$ is the bulk scalar field
potential, while $\mathcal{V}_{1}(\psi^{1})$ and
$\mathcal{V}_{2}(\psi^{2})$ are boundary potentials. Finally, for
the matter fields confined to the branes, we shall consider the
standard action:
\begin{eqnarray}
S_{\mathrm{BR}} = S_{1}[\Psi_{1},g_{\mu \nu}(z_{1})] +
S_{2}[\Psi_{2},g_{\mu \nu}(z_{2})],
\end{eqnarray}
where $\Psi_{1}$ and $\Psi_{2}$ denote the respective matter
fields, and $g_{\mu \nu} (z_{a})$ is the induced metric at
position $z_{a}$. In the present case (BPS-configurations), we
shall consider the following general form for the potentials:
$\mathcal{U} = U + u$, $\mathcal{V}_{1} = U_{B} + v_{1}$ and
$\mathcal{V}_{2} = U_{B} + v_{2}$,
where $U$ and $U_{B}$ are the bulk and brane superpotentials, and
the potentials $u$, $v_{1}$ and $v_{2}$ are such that \mbox{$|u|
\ll |U|$} and $|v_{1}|, |v_{2}| \ll |U_{B}|$. In this way, the
system is dominated by the superpotentials $U$ and $U_{B}$.  The
most important characteristic of this class of system is the
relation between $U$ and $U_{B}$ (the BPS-relation), given by:
\begin{eqnarray} \label{eq: BPS-cond}
U= \left( \partial_{\psi} U_{B}  \right)^{2} - U_{B}^{2}.
\end{eqnarray}
This specific configuration, when the potentials $u, v_{1}, v_{2}
= 0$ and no fields other than the bulk scalar field are present,
is the BPS configuration. When $U_{B}$ is the constant
potential, the Randall--Sundrum model is recovered with a bulk
cosmological constant $\Lambda_{5} = (3/8) U =  - (3/8) \,
U_{B}^{2}$. The presence of the potentials $u$, $v_{1}$ and
$v_{2}$ are generally expected from supersymmetry breaking
effects.


To develop the moduli space approximation of the present system,
we need to know its static vacuum solution. To this extent,
consider a bulk scalar field $\psi_{0}$ and gravitational fields
$N_{0}$ and $\tilde g_{\mu \nu}(x)$, such that:
\begin{eqnarray}
\partial_{\mu} \psi_{0} = 0, \,\, \quad
\partial_{\mu} N_{0} = 0, \,\, \quad \mathrm{and} \,\, \quad
\tilde G_{\mu \nu} = 0,
\end{eqnarray}
(where $\tilde G_{\mu \nu}$ is the four-dimensional Einstein
tensor constructed from $\tilde g_{\mu \nu}$) and also consider a
metric $g_{\mu \nu} = \omega^{2}(z) \tilde g_{\mu \nu}$ so that
the $z$ dependence of $g_{\mu \nu}$ is only contained in the warp
factor $\omega (z)$. Then, let us assume that these fields satisfy
the following two relations:
\begin{eqnarray}
\frac{\omega_{0}'}{\omega_{0}} = - \frac{1}{4} N_{0} U_{B}, \quad
\quad \psi_{0}' = N_{0} \frac{ \partial U_{B}}{\partial \psi_{0}}.
\label{eq: match}
\end{eqnarray}
These are the BPS relations, which agree with the
boundary conditions for the bulk fields in the absence of matter
and supersymmetry breaking terms. On the other hand, when they hold,
then $\psi_{0}$, $N_{0}$ and
$g_{\mu \nu} = \omega^{2}(z) \tilde g_{\mu \nu}$ solve the entire
system of equations of motion. This important fact constitutes one
of the main properties of BPS-systems, and means that the branes
can be arbitrarily located anywhere in the background, without
obstruction. It should be clear, though, that when matter is allowed to
exist in the branes the boundary conditions will not continue
being solutions to the equations of motion, and the static
configuration will not be possible; the presence of
matter in the branes drives the system to a cosmological
evolution.

In the static vacuum solution expressed through the equations in
(\ref{eq: match}), the dependence of the lapse function $N_{0}$, in
terms of $z$ is completely
arbitrary (though it must be restricted to be positive)
and its precise form will correspond to a gauge
choice. Let us assume that $\psi_{0}(z)$ has boundary values:
\begin{eqnarray}
\psi^{1} = \psi_{0}(z_{1}), \quad \mathrm{and} \quad \psi^{2} =
\psi_{0}(z_{2}).
\end{eqnarray}
Since we are interested in the static vacuum solution, $\psi^{1}$
and $\psi^{2}$ are just constants. The precise form of
$\psi_{0}(z)$, as a function of $z$, depends on the form of
$U_{B}(\psi)$ and the gauge choice for $N_{0}$. However, it is not
difficult to see that $\psi^{1}$ and $\psi^{2}$ are the only
degrees of freedom, jointly with $\tilde g_{\mu \nu}$, necessary
to specify the BPS state of the system. That is, given a gauge
choice $N_{0}$, we have: $\psi_{0} = \psi_{0}(z, \psi^{1}, \psi^{2})$ and
$N_{0} = N_{0}(z, \psi^{1}, \psi^{2})$.
Moreover, it is possible to show that by
virtue of the relations in (\ref{eq: match}), the solution
$\psi_{0}$ must be monotonic in term of $z$ in the complete bulk
space \cite{Palma-Davis}. Therefore, the change of
variable $dz = N_{0}^{-1} \left(
\partial_{\psi_{0}} U_{B} \right)^{-1} \! d \psi_{0}$ can be used to
parameterize the fifth dimension in terms of $\psi_{0}$, and
the boundary values $\psi^{1}$ and $\psi^{2}$ specify the positions of the
branes.

Observe additionally, from eq. (\ref{eq: match}), that $\omega_{0}(z)$ can be expressed in
terms of $\psi_{0}(z)$ in a gauge independent way:
\begin{eqnarray}
\omega_{0}(z) &=& \exp \left[ -\frac{1}{4}
\int_{\psi^{1}}^{\psi_{0}(z)}
\!\! \alpha^{-1}(\psi) \, d \psi \right], \label{eq: omega} \\
\alpha(\psi) &=& \frac{1}{U_{B}} \frac{\partial U_{B}}{\partial
\psi}.
\end{eqnarray}
In the last equations we have normalized the solution
$\omega_{0}(z)$ in such a way that the induced metric to the first
brane is $\tilde g_{\mu \nu}$. The induced metric on the second
brane is, therefore, conformally related to the first brane, with
a warp factor $\omega_{0}(z_{2})$.


We now proceed to compute the moduli-space approximation. First of
all, varying the action $S_{\mathrm{tot}}$ in terms of $N$, we
deduce the following equation of motion:
\begin{eqnarray}
K^{2} -  K_{\mu \nu} K^{\mu \nu} = R  +
  \frac{3}{4} \bigg[ \frac{1}{N^{2}}
(\psi')^{2} - (\partial \psi)^{2} - \mathcal{U} \bigg]. \label{eq:
Grav-eq}
\end{eqnarray}
This result can be inserted back into the action (\ref{eq:
Tot-Act}):
\begin{eqnarray} \label{eq: S-tot 2}
S_{\mathrm{tot}} &=& \frac{1}{\kappa_{5}^{2}} \int_{M} \!\! dz \,
d^{\, 4} x \sqrt{-g} \, N \Big[ R - \frac{3}{4} (\partial
\psi)^{2} - \frac{3}{4} \mathcal{U} \Big] +
S^{1}_{\psi} + S^{2}_{\psi} + S_{\mathrm{BR}}.
\end{eqnarray}
We can now exploit the static vacuum solution. As we mentioned,
when matter is present in the branes as well as supersymmetry
breaking terms, the system becomes dynamical.
Hence, the boundary fields $\psi^{1}$ and $\psi^{2}$ and
the metric $\tilde g_{\mu \nu}$ no longer satisfy vacuum equations
of motion; instead we insert $\psi_{0}$, $N_{0}$ and $g_{\mu \nu} =
\omega^{2} \tilde g_{\mu \nu}$ as ansatzs into the action (\ref{eq:
S-tot 2}). That is, the positions of the branes, which are parameterized
by the moduli $\psi^{1}$ and $\psi^{2}$,  are being perturbed by
the matter content of the branes. Thus we obtain the following
action for $\psi^{1}$, $\psi^{2}$ and $\tilde g_{\mu \nu}$:
\begin{eqnarray} \label{eq: S-tot 3}
S_{\mathrm{tot}} &=& \frac{1}{\kappa_{5}^{2}} \int_{M} \!\! dz
\, d^{\, 4} x \sqrt{-\tilde g} \, N_{0} \omega_{0}^{4} \Big[
\omega_{0}^{-2} \tilde R  - 6 \omega_{0}^{-3} \Box \omega_{0} - \frac{3}{4} \omega_{0}^{-2} (\partial
\psi_{0})^{2} - \frac{3}{4} u \Big] \nonumber \\ && - \frac{3}{2 \kappa_{5}^{2}}
\int_{\sigma_{1}} \!\!\! d^{4} x \sqrt{-g} \, v_{1}  + \frac{3}{2 \kappa_{5}^{2}} \int_{\sigma_{2}} \!\!\! d^{4} x
\sqrt{-g} \, v_{2} + S_{\mathrm{BR}},
\end{eqnarray}
where the BPS relation (\ref{eq: match}) was used to evaluate some
terms at the boundaries. To obtain a more conventional form for the
action in terms of $\psi^{1}$, $\psi^{2}$ and $\tilde g_{\mu
\nu}$ we need to integrate along the fifth dimension. To do this
it is necessary to note the following identity from eq. (\ref{eq: match}):
\begin{eqnarray}
\partial_{\mu} (N_{0} \omega_{0}) = -N_{0} \alpha_{0} \omega_{0}
\partial_{\mu} \psi_{0} - \frac{4}{U_{B}} ( \partial_{\mu}
\omega_{0})'.
\end{eqnarray}
This expression allows us to rewrite the term $ 6 (\partial
\omega_{0} )^{2} + (3/4)(\partial \psi_{0})^{2}$, present in the
action (\ref{eq: S-tot 3}), as:
\begin{eqnarray}
6 \omega_{0}^{-1} \Box \omega_{0} + \frac{3}{4} (\partial
\psi_{0})^{2} = 12 (N_{0} \omega_{0}^{4})^{-1} \left[
\frac{1}{U_{B}} (\partial \omega_{0})^{2} \right]' + \frac{3}{4} \omega_{0}^{-2} \alpha_{0}^{2} \alpha_{1}^{-2}
(\partial \psi^{1})^{2},
\end{eqnarray}
where $\alpha_{0} = \alpha(\psi_{0})$ and $\alpha_{1} =
\alpha(\psi^{1})$. Now, using the parameterization $dz = N_{0}^{-1} \left(
\partial_{\psi_{0}} U_{B} \right)^{-1} \! d \psi_{0}$ to integrate along
the fifth dimension and the boundary values
$\psi^{1}(x)$ and $\psi^{2}(x)$ to evaluate at the
positions of the branes $z_{1}$ and $z_{2}$, we arrive at the
following effective theory:
\begin{eqnarray}
S &=& \frac{1}{k \kappa_{5}^{2}} \int d^{4}x \sqrt{- \tilde g}
\bigg[ \Omega^{2} \tilde R - \frac{3}{4} \tilde g^{\mu \nu}
\gamma_{a b}
\partial_{\mu} \psi^{a} \partial_{\nu} \psi^{b}
- \frac{3}{4} V  \bigg]  \nonumber\\
&& + S_{1}[\Psi_{1}, \tilde g_{\mu \nu}] + S_{2}[\Psi_{2},
\omega^{2}(z_{2}) \tilde g_{\mu \nu}], \qquad \label{eq5: EFF
Action}
\end{eqnarray}
where the index $a$ labels the positions $1$ and $2$.
The conformal factor $\Omega^{2}$ in front of the Ricci
scalar $\tilde R$, is given by:
\begin{eqnarray}
\Omega^{2} = k \int_{\psi^{1}}^{\psi^{2}} \!\!\!\!\! d\psi \left(
\frac{\partial U_{B}}{\partial \psi} \right)^{-1} \!\!\!
\omega^{2}, \label{eq5: Big-Omega}
\end{eqnarray}
where $\omega$ is given by equation (\ref{eq: omega}). The
coefficient $k$ is an arbitrary positive constant with dimensions
of inverse length to make
$\Omega^{2}$ dimensionless. The symmetric matrix $\gamma_{a b}$ depends on
 the moduli fields, and can be regarded as the
metric of the moduli space in a sigma model
approach. Additionally, the elements of $\gamma_{a b}$ are given
by:
\begin{eqnarray}
\gamma_{1 1} &=& \alpha_{1}^{-2} \left[ \frac{k}{U_{B}(\psi^{1})}
- \frac{1}{2} \Omega^{2}  \right],  \\
\gamma_{2 2} &=& \alpha_{2}^{-2}  \frac{
\omega^{2}(z_{2}) k}{U_{B}(\psi^{2})}, \\
\gamma_{1 2} &=& - \alpha_{1}^{-1} \alpha_{2}^{-1}
\frac{\omega^{2}(z_{2}) k}{U_{B}(\psi^{2})},
\end{eqnarray}
with $\gamma_{2 1} = \gamma_{1 2}$. Finally, we have also defined
an effective potential $V$ which depends linearly on the
supersymmetry breaking potentials $u$, $v_{1}$ and $v_{2}$. This
is defined as:
\begin{eqnarray}
V = \frac{k}{2} \int^{\psi^{2}}_{\psi^{1}} \!\!\!\!\! d\psi \left(
\frac{\partial U_{B}}{\partial \psi} \right)^{-1} \!\!\!
\omega^{4} \, u - 2 k \left[ \omega^{4}(z_{2}) v_{2} - v_{1}
\right].
\end{eqnarray}
The generic form of the deduced theory is of a bi-scalar tensor
theory of gravity, with the two scalar degrees given by $\psi^{1}$
and $\psi^{2}$. Note that in equation (\ref{eq5: EFF Action}) the
Newton's constant depends on the moduli fields. This theory can be
rewritten in the Einstein frame where the Newton's constant is
independent of the moduli. By considering the conformal
transformation  $\tilde g_{\mu \nu} \rightarrow g_{\mu \nu} =
\Omega^{2} \tilde g_{\mu \nu}$ we are then left with the following
action:
\begin{eqnarray}
S &=& \frac{1}{k \kappa_{5}^{2}} \int d^{4}x \sqrt{- g} \bigg[ R -
\frac{3}{4} g^{\mu \nu} \gamma_{a b}
\partial_{\mu} \psi^{a} \partial_{\nu} \psi^{b} - \frac{3}{4} V \bigg]  \nonumber\\
&& + S_{1}[\Psi_{1}, A_{1}^{2} g_{\mu \nu}] + S_{2}[\Psi_{2},
A_{2}^{2} g_{\mu \nu}], \label{eq: EFF Action}
\end{eqnarray}
where now the sigma model metric $\gamma_{a b}$ is given by:
\begin{eqnarray}
\gamma_{1 1} &=&  2 \alpha_{1}^{-2} \frac{k^{2} A_{1}^{4}}
{U_{B}^{2}(\psi^{1})} \bigg[ 1 - \frac{1}{2 k} U_{B}(\psi^{1})
A_{1}^{-2}
\bigg] , \\
\gamma_{2 2} &=& 2 \alpha_{2}^{-2} \frac{k^{2} A_{2}^{4}}
{U_{B}^{2}(\psi^{2})} \bigg[ 1 + \frac{1}{2 k} U_{B}(\psi^{2})
A_{2}^{-2}
\bigg]  , \\
\gamma_{1 2} &=& - 2 \alpha_{1} \alpha_{2} \frac{ k^{2} A_{1}^{2}
A_{2}^{2} } {U_{B}(\psi^{1})  U_{B}(\psi^{2}) }.
\end{eqnarray}
It is possible to show that in this frame $\gamma_{a b}$ is a
positive definite metric. Additionally, we have defined the
quantities $A_{1}$ and $A_{2}$ (which are functions of the moduli)
to be $A_{1}^{-2} = \Omega^{2}$ and $A_{2}^{-2} = \Omega^{2}
\omega^{-2}(z_{2})$, or explicitly:
\begin{eqnarray}
A_{1}^{-2} = k \!\! \int_{\psi^{1}}^{\psi^{2}} \!\!\!\!\! d\psi
\left( \frac{\partial U_{B}}{\partial \psi} \right)^{\!\! -1}
\!\!\!\!\! \exp \left[ -\frac{1}{2} \int_{\psi^{1}}^{\psi}
\!\!\!\! \alpha^{-1} d \psi \right], \qquad \\
A_{2}^{-2} = k \!\! \int_{\psi^{1}}^{\psi^{2}} \!\!\!\!\! d\psi
\left( \frac{\partial U_{B}}{\partial \psi} \right)^{\!\! -1}
\!\!\!\!\! \exp \left[ -\frac{1}{2} \int_{\psi^{2}}^{\psi}
\!\!\!\! \alpha^{-1} d \psi \right]. \qquad
\end{eqnarray}
Also, the potential $V$ is now found to be:
\begin{eqnarray}
V &=& \frac{k}{2} \Omega^{-4} \int^{\psi^{2}}_{\psi^{1}}
\!\!\!\!\! d\psi \left( \frac{\partial U_{B}}{\partial \psi}
\right)^{-1} \!\!\! \omega^{4} \, u - 2 k \left[ A_{2}^{4} v_{2} -
A_{1}^{4} v_{1} \right]. \label{eq: Pot V}
\end{eqnarray}
Our effective action can be used extensively for the study of this
class of systems. It can also be obtained in the projective
approach, when a perturbative method is used to analyze the
five-dimensional equations of motion \cite{Palma-Davis}.
Additionally, it agrees with previous computations, where the
specific case of dilatonic branes, $U_{B} \propto e^{\alpha
\psi}$, was considered \cite{Br-CvdB-ACD-Rh}. To obtain the next
order in the moduli space approximation, we should consider linear
perturbations about the vacuum solution. That is, we should
consider: $g_{\mu \nu} = \omega_{0}^{2} \left( \tilde g_{\mu \nu}
+ h_{\mu \nu} \right)$, $\psi = \psi_{0} + \varphi$ and  $N =
N_{0} \, e^{\phi}$, with the linear fields satisfying $|h_{\mu
\nu}| \ll | \tilde g_{\mu \nu}|$, $|\varphi| \ll |\psi_{0}|$ and
$|\phi| \ll 1$. The study of these linear perturbation was
considered in detail in \cite{Palma-Davis}.

When the cosmological evolution of branes is considered it is
possible show that the branes are driven by the matter content in
them. In particular, the first brane, $\sigma_{1}$, is driven
towards the minimum of the BPS potential, while the second brane,
$\sigma_{2}$, is driven towards the maximum \cite{Palma-Davis}.
This allows the system to fall in a
stable configuration. For example, we can compute the Post
Newtonian Eddington coefficient $\gamma$ which is constrained by
measurements of the deflection of radio waves by the Sun to be $ \gamma = 1
+ (2.1 \pm 2.3) \times 10^{-5}$
\cite{Eddington}. The parameter $\gamma$ is found to be:
\begin{eqnarray}
1 - \gamma \simeq  \frac{2}{3} \left[ 1  - \frac{1}{2 k}
U_{B}(\psi^{1}) A_{1}^{-2} \right]. \label{eq: gamma-edd}
\end{eqnarray}
This is a very important result: since the branes must be
near the extremes of the BPS potential, observational measurements
constrain the global configuration of the brane system, as well as
the shape of the potential.

Summarizing, in this paper we have developed the moduli-space
approximation of the low energy regime of BPS brane-world models.
As a result, an effective 4-dimensional system of equations have
been obtained. At this order, the metrics of both branes are
conformally related, and the complete theory corresponds to a
bi-scalar tensor theory of gravity [equation (\ref{eq: EFF
Action})]. Our effective theory allows the study of this class of
models within the approach and usual techniques of multi-scalar tensor
theories \cite{Damour}. For instance, we have indicated that the moduli
fields can be stabilized, and that the system can be
constrained by observations. \\

We are grateful to Philippe Brax and Carsten van de Bruck for
useful comments and discussions. This work is supported in part by
PPARC and MIDEPLAN (GAP).

\end{document}